# Data-driven, metaheuristic-based off-grid microgrid capacity planning optimisation and scenario analysis: Insights from a case study of Aotea–Great Barrier Island

Soheil Mohseni, Roomana Khalid, and Alan C. Brent

Sustainable Energy Systems, Wellington Faculty of Engineering, Victoria University of Wellington, Wellington, 6140, New Zealand

Presenter:

Alan C. Brent

**Abstract**

Given their relatively small markets, vulnerable energy assets, and the need for fuel imports (in diesel-dependent communities), remote coastal and island communities are often faced with higher levelised costs of energy compared to those that have access to national grids. Accordingly, small privately-purchased off-grid renewable energy systems (RESs) are increasingly used for energy generation in these areas. However, such privately-purchased stand-alone RESs are often unaffordable for households with lower incomes. The household-level isolated RESs additionally suffer from the lack of onsite capacity for the optimal operation and maintenance of the installed equipment. These bring to light the importance of information and communications technologies-enabled, remotely-operated islanded microgrids for the more optimal coordination of integrated distributed energy resources – including renewable energy generation and storage technologies. This, accordingly, paves the way for the provision of reliable, resilient, affordable, clean energy in remote areas – particularly using community-based business models. To this end, advanced off-grid microgrid capacity planning optimisation models are needed. In this context, while considerable attention has been devoted to a range of off-grid microgrid sizing methods, leveraging the potential of data-driven, artificial intelligence-based metaheuristic optimisation algorithms is less well-explored. Importantly, data-driven metaheuristics have the potential to produce the nearest solution to the globally optimum solution in microgrid sizing applications, which have been recognised as non-deterministic, polynomial time-hard (NP-hard) problems. Furthermore, there is a general lack of electrified transportation interventions considered during long-term grid-independent microgrid planning phases. In response, this paper introduces a novel metaheuristic-based strategic off-grid microgrid capacity planning optimisation model that is applicable to associated integrated energy and e-mobility resource plans. The formulated general off-grid microgrid sizing model is solved using a competitively selected state-of-the-art metaheuristic, namely moth-flame optimisation. To test the effectiveness of the proposed model, three independent microgrid development projects have been considered for three communities residing on Aotea–Great Barrier Island – Tryphena, Medlands, and Mulberry Grove. The sites of interest have different demand profiles and renewable energy potentials – with consequent changes in the technologies considered in the associate candidate pools. Moreover, to generate key insights into the impact of different variables on the costing and configuration of the associated microgrids, a number of scenario analyses have been performed. Importantly, comprehensive numeric simulation results demonstrate the economic viability of the project proposals formulated based on the proposed microgrid planning method.

## 1. Introduction

Providing electricity to the estimated 1 billion people remaining without access globally has been found to be particularly challenging given the costs associated with the expansion of conventional power networks to remote areas with geographically challenging terrain [1]. This brings to light the importance of distributed energy generation technologies that are sited close to the point of consumption, such as wind turbines (WTs), solar photovoltaic (PV) panels, micro-hydropower plants (MHPPs), and so forth. It is increasingly recognised that distributed energy generation provides an effective platform for the electrification of remote areas, as well as decarbonising power supply [2], [3].

In this context, the microgrid (MG) concept has emerged to integrate several clean energy micro-sources into a dispatchable (controllable) system with a prescribed reliability level. The U.S. Department of Energy defines a MG as "*a group of interconnected loads and distributed energy resources within clearly defined electrical boundaries that acts as a single controllable entity with respect to the grid*" [4]. The flexible architecture of MGs allows for the optimal integration of distributed renewable energy sources (RESs) into the system – towards increasing the penetration of renewables in both urban regions and remote, rural localities [5].

The optimal sizing of the components of MGs is necessary to ensure a cost-minimal power supply whilst adhering to a set of operational and planning constraints. Accordingly, many studies in the literature have focused on developing MG capacity planning optimisation methods. A comprehensive review of existing MG planning methods can be found in [6], which indicates that analytical approaches are the mainstream optimisation strategy. The major issues associated with analytical approaches to MG sizing are the strong assumptions and simplifications involved in such approaches, which lead to significant simulation-to-reality gaps, especially in terms of the optimal cost of the system [7].

Alternatively, meta-heuristic algorithms are increasingly utilised to approximate solutions to energy planning optimisation problems due to their applicability to the original (unreduced) problems. For instance, Bilal et al. [8] have calculated the optimal size of WTs, PV panels, and battery packs in a MG based on cost minimisation using a genetic algorithm-based solution approach. Radosavljević et al. [9] have also proposed optimising the energy and operational management of grid-connected MGs using the particle swarm optimisation (PSO) technique. However, less attention has been given to optimising the size of MGs using newly developed meta-heuristic algorithms such as the moth-flame optimisation algorithm (MFOA) [10]. Accordingly, the extent to which these algorithms might outperform the well-established algorithms is not yet well known. Also, the integration of transport electrification interventions in the integrated resource planning optimisation processes of off-grid MGs is less well-explored. This paper, therefore, seeks to address these knowledge gaps by introducing a novel meta-heuristic-based MG capacity planning optimisation method tailored specifically to stand-alone applications in the presence of a significant volume of EV-charging loads.

It is also noteworthy that Aotearoa–New Zealand aims to achieve net-zero carbon emissions by 2050. Currently, around 40% of primary energy supply and more than 75% of electricity demand is met by RESs in Aotearoa–New Zealand. Also, Aotearoa–New Zealand holds third place in the Organisation for Economic Cooperation and Development (OECD) in terms of the share of renewable electricity generation [11]. This makes case studies in Aotearoa–New Zealand well-suited candidates for evaluating the effectiveness of new MG sizing methods.

## 2. Mathematical modelling of test-case microgrids

To test the effectiveness of the proposed optimisation model, two off-grid MG structures are considered. Collectively, the components of the proposed MGs include solar photovoltaic (PV) panels, wind turbines (WTs), battery storage systems (BSSs), and various power converters. More specifically, MG 1 is driven by wind and solar PV resources, whereas MG 2 is assumed to be solely driven by solar PV power. The employed power conversion apparatuses in the two MGs can be classified as DC/AC inverters, AC/DC converters, and DC/DC converters.

The schematic diagram of the first stand-alone MG system is depicted in Fig. 1. It uses solar PV and WT technologies for power generation, which are supported by a BSS. A dump load is also considered to be able to maintain the balance of power supply and demand when total non-dispatchable generation outstrips total loads. Furthermore, the optimal size of all power conditioning devices that lie between the generation/storage components and the DC bus is assumed to be equal to the optimal size of the devices they couple to the common DC bus. Moreover, as it can be seen from the figure, the total load on the system can be broken down into residential, commercial, and electric vehicle (EV)-charging loads. Accordingly, the optimal capacities of the residential and commercial loads' inverters, as well as the EV-charging loads' inverters, are controlled by the relevant peak demands. The following sections mathematically model the components of MG 1 and define the operational strategy of the network. The model of MG 2 is similar except that it does not incorporate WTs in the candidate pool.

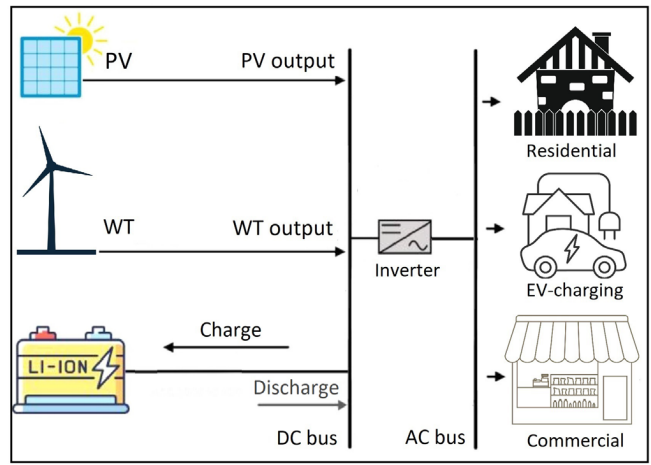

Figure 1. Schematic diagram and energy flow of MG 1.

*2.1. Wind turbines*

The 3-blade Senwei SWT-50 kW wind turbine is selected, which has a rated power of 50 kW AC with a hub altitude of 34 m. The wind speed data, measured at the height of $h_{ref}$, is normalised to the hub height $h$ using the following equation [12]:

$$V_h = V_{ref} \times (h/h_{ref})^r, \qquad (1)$$

where $V_{ref}$ is the reference wind speed recorded at the height of $h_{ref}$, $r$ is an exponent in the range [0.1, 0.25], which reflects the characteristics of the terrain. Given the non-flat, tree-covered land characteristics of the site, $r$ is considered to be 0.2 in this study.

Stewart and Essenwanger [13] have derived the power output from WTs as a more flexible version of the general Weibull distribution of WT power output, which can be expressed as:

$$P_{WT}(t) = N_{WT} \times \begin{cases} 0; & if \ v < v_{cin} \ or \ v > v_{cout} \\ P_{rated} \times \left(\dfrac{v(t)^3 - v_{cin}^3}{v_{rated}^3 - v_{cin}^3}\right); & if \ v_{cin} \le v < v_{rated} \\ P_{rated}; & if \ v_{rated} \le v \le v_{cout} \end{cases} \qquad (2)$$

where $P_{WT}$, $P_{rated}$, $v(t)$, $v_{cin}$, and $v_{cout}$ respectively denote the power output from the WT (kW), the turbine's rated power (kW), normalised wind speed (m/s) at time-step $t$, and the associated cut-in and cut-out wind speeds (kW), and $N_{WT}$ is the optimum number of WTs.

### 2.2. PV panels

In this study, the Half Cut PERC Mono Photovoltaic solar panel is considered. The panel has a rated capacity of 330 W with an expected lifetime of 25 years. Eq. (3) can be used to calculate the aggregate power output from the solar generation system at time-step $t$ [14]:

$$P_{PV}(t) = N_{PV} \times DF \times G_t(t)/1000, \qquad (3)$$

where $N_{PV}$ is the optimum number of solar panels, which is updated at each iteration of the sizing process, and $G_t(t)$ is the global solar irradiance ($W/m^2$) at time-step $t$. A calendar-induced degradation factor (DF) of 85% was considered for the solar PV generation system. Also, a tilt angle of 36° was considered for the PV panels.

### 2.3. Battery storage

The 14-kWh Tesla Powerwall battery pack is considered in this study. The following equality constraint is used to ensure that the battery bank's state of charge (SOC) at each time-step $t$ is aware of the underlying charging and discharging processes [15]:

$$E_b(t) = E_b(t-1) + (P_{ch}(t) \times \eta_{ch}) \times \Delta t - (P_{dch}(t)/\eta_{dch}) \times \Delta t, \qquad (4)$$

where $P_{ch}$ is the battery bank's charging power, $P_{dch}$ is the battery bank's discharging power, while $\eta_{ch}$ and $\eta_{dch}$ are the charging and discharging efficiencies of the battery packs, respectively, both of which are assumed to be 95% in this study.

### 2.4. Summary of techno-economic specifications

Table 1 summarises the techno-economic specifications of the selected components in the candidate pool for MG 1. It is noteworthy that all costs are cited in 2021 New Zealand Dollars.

Table 1. Techno-economic specifications of the candidate components of MG 1.

| Component | Manufacturer part number | Nameplate rating | Capital cost | Replacement cost | O&M cost | Lifetime (years) |
|---|---|---|---|---|---|---|
| Wind turbine | Senwei SWT | 50 kW | $65,000/unit | $65,000/unit | $900/unit/year | 25 |
| Solar PV panel | Half Cut PERC | 0.33 kW | $335/unit | $335/unit | $6/unit/year | 25 |
| Battery pack | Powerwall | 14 kWh | $15,000/unit | $11,000/unit | $30/unit/year | 15 |
| AC/DC conv. | Red Prime | 50 kW | $4,770/unit | $4,770/unit | $28/unit/year | 15 |
| DC/DC conv. | SAMLEX IDC | 0.36 kW | $50/unit | $50/unit | $1/unit/year | 15 |
| Inverter | Eaton DG IP21 | 21 kW | $8,000/unit | $8,000/unit | $320/unit/year | 20 |
| EV-charger | SolarEdge | 7.6 kW | $4,000/unit | $4,000/unit | $160/unit/year | 20 |

*2.5. Dispatch strategy*

A rule-based, cycle-charging energy management strategy is specifically developed to ensure that generation always meets demand. To this end, the battery storage linking variables are used. Two underlying dispatch strategies tailored to different scenarios are considered for the operation. More specifically, when the aggregate power output from renewable energy generation technologies is greater than the total load demand (including the EV-charging loads), the excess energy generation is stored in the BSS. On the other hand, when the total residential/commercial load demand is greater than the total power output from renewable energy generation technologies, the battery bank is discharged to meet the residential and commercial loads as far as possible. However, for reasons of energy efficiency, the EV-charging load is not served by discharging the stationary battery bank, which might lead to the loss of total EV-charging loads, in addition to potentially part of the residential and commercial loads.

## 3. Meta-heuristic-based microgrid capacity planning optimisation

The objective of the proposed MG life-cycle cost estimation method is to minimise the whole-life cost of off-grid MGs based on net present valuations subject to a set of operational- and planning-level constraints. More specifically, the objective function consists mainly of the size of the equipment multiplied by the associated per-unit cost factors, as [16]:

$$NPC_c = N_c \times \left( C_c + RC \times SPPW + \frac{C_{O\&M}}{CRF(i,T)} - SV \right), \tag{5}$$

where $NPC_c$ is the net present cost of component $c$, while $N_c$, $C_c$, $C_{O\&M}$, and $RC$ represent the optimal capacity, capital cost, operation and maintenance cost, and replacement cost of the MG component $c$, respectively. Also, $SPPW$, $CRF$ and $SV$ respectively denote the single payment present worth, capital recovery factor, and salvage value of the corresponding component.

The salvage value can be determined as follows:

$$SV = RC \times \frac{L - (T - L \times [\frac{T}{L}])}{L}, \tag{6}$$

where $L$ and $T$ respectively denote the expected lifetime of the associated component (years) and the expected lifetime of the MG system (years), which is assumed to be 25 years.

Furthermore, $SPPW$ represents the present value of a one-time cash outflow corresponding to a series of equal future payments, which can be modelled as:

$$SPPW = \sum_{n=1}^{Y} \frac{1}{(1+i)^{L \times n}}, \tag{7}$$

where $Y = \left[\frac{T}{L}\right]$, and $i$ is the real interest rate, which is assumed to be 6% in this study.

Moreover, the capital recovery factor is the ratio of a constant annuity to the corresponding present value for a considered length of time, which can be calculated as follows:

$$CRF(i,t) = \frac{i(1+i)^T}{(1+i)^T - 1}. \tag{8}$$

Accordingly, the total net present costs (TNPCs) of MGs 1 and 2 can be modelled as:

$$TNPC_{MG1} = NPC_{PV} + NPC_{WT} + NPC_{BSS} + NPC_{Conv} + NPC_{EV-ch} + pen1 + pen2, \tag{9}$$

$$TNPC_{MG2} = NPC_{PV} + NPC_{BSS} + NPC_{Conv} + NPC_{EV-ch} + pen1 + pen2, \quad (10)$$

where $NPC_{PV}$, $NPC_{WT}$, $NPC_{BSS}$, $NPC_{Conv}$, and $NPC_{EV-ch}$ respectively denote the net present costs of the PV panels, WTs, the BSS, the converters, and EV-chargers. Also, $pen1$ and $pen2$ are the penalty terms (as sufficiently large values) associated with not meeting the desired reliability levels in serving the residential/commercial and EV-charging loads, respectively. That is, they are employed to mark the positions in the search space where any of the imposed constraints are violated as infeasible equipment size combinations.

The equivalent loss factor (ELF) is used to measure the reliability of the system, which, unlike other relevant reliability indicators, is aware of both the frequency and magnitude of lost loads. In the context of this study, the ELF associated with unmet residential/commercial and EV-charging loads can be expressed as:

$$ELF_{load} = \frac{1}{n}\sum_{t=1}^{n} \frac{Q_{load}(t)}{P_{load}(t)}, \quad (11)$$

$$ELF_{EV} = \frac{1}{n}\sum_{t=1}^{n} \frac{Q_{EV}(t)}{P_{EV}(t)}, \quad (12)$$

where $n$ is the number of time-steps in the planning horizon, $ELF_{load}$ is associated with the residential/commercial loads (assumed to be 0), while $ELF_{EV}$ is associated with EV-charging loads (assumed to be 0.005).

To optimise the formulated problem, the moth-flame optimisation algorithm (MFOA) was chosen. It is a state-of-the-art meta-heuristic optimisation algorithm, the superiority of which to a wide range of meta-heuristics has been statistically demonstrated in the literature [17]–[19].

## 4. Case studies

Aotea–Great Barrier Island is situated in the outer Hauraki Gulf, 100 km north-east of central Auckland, with an area of 285 square kilometres and the following coordinates: latitude 36.26˚S and longitude 175.49˚E. It is the sixth-largest island of Aotearoa–New Zealand. According to the 2018 census, the island has a usually resident population of 936 people [20]. However, the total population of the island increases significantly over holiday periods due to tourism.

*4.1. Climatic conditions of Aotea–Great Barrier Island*

Aotea–Great Barrier Island is richly endowed with a solar resource in summer, with a daily average of 11.3 hours of sunlight during the summertime according to the SolarView database of NIWA [21]. More specifically, December is the sunniest month – with 6.4 hours of sunlight per day (on average) – whereas May is associated with the least sunlight hours – with an average of 3.3 hours of sunlight per day.

In terms of wind resources, October is the windiest month, with an average maximum wind speed of approximately 34 km/h, whereas April is the weakest month in terms of wind resources, with an average minimum wind speed of around 14 km/h. The solar and wind resources exhibit significant complementary characteristics. Prior techno-economic feasibility assessments have estimated that harnessing solar and wind resources for electricity generation is able to serve the energy needs of the island's inhabitants in a reliable, affordable, self-sufficient, sustainable, and socially acceptable manner [22].

As mentioned above, the solar irradiance data were retrieved from the NIWA's SolarView database [21], while the wind speed data were collected from the CliFlo database of NIWA [23].

To this end, 15 years' worth (2007 to 2021) of hourly-basis, year-round historical solar irradiance and wind speed records (8,760 data points) were collected from the relevant databases for the three micro-communities residing on the island, namely: Medlands (MG 1), Tryphena (MG 2a), and Mulberry Grove (MG 2b).

*4.2. Meteorological data*

The meteorological data requirements of the three MGs are as follows: solar irradiance (MGs 1, 2a, 2b), and wind speed (MG 1). It should be recalled that MG 1 integrates both solar PV and WT technologies, whereas MGs 2a and 2b are driven solely by solar PV resources. That is, prior techno-economic feasibility and business case analyses have indicated that the WT technology is a non-viable choice for Tryphena and Melburry Grove given the hilly, tree-covered terrain. Fig. 2 shows the geographical locations of the three communities considered for off-grid MG installations. In terms of geographical location, MGs 2a and 2b are about 6.3 km and 8.1 km far from MG 1, respectively.

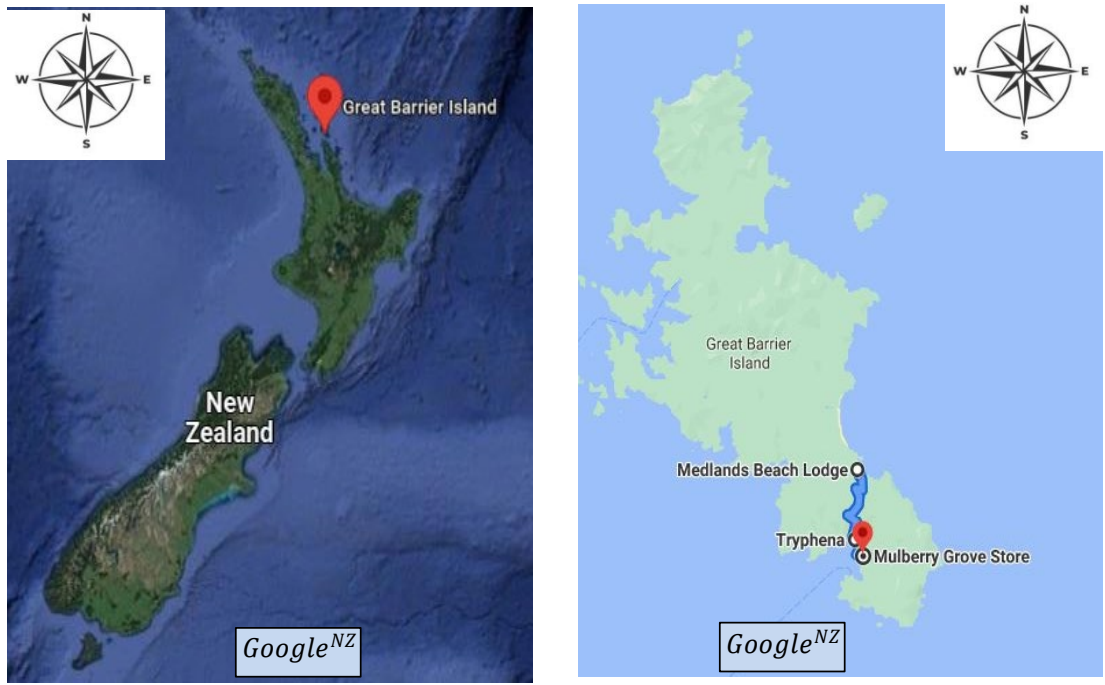

Figure 2. Locations of the conceptualised MGs for installation on Aotea–Great Barrier Island (image courtesy of Google Earth™ mapping service).

*4.3. Load data*

Given the lack of reliable historical data, residential load demand was synthesised based on typical residential load profiles [17]. Table 2 presents the number of residential and commercial end-use buildings in the cases of interest, which are used for synthetic load profiling. As the table details, a total of 43 buildings were considered for the case of MG 1, while the total number of buildings connected to MG 2a and MG 2b were 36 and 18, respectively. The reason for a relatively higher number of buildings in MG 1 compared to MGs 2a and 2b is that it is recognised as the island's most popular location among visitors coming for holidays or leisure [22].

Table 2. Number of buildings in the three MGs.

| Microgrid | MG 1 | MG 2a | MG 2b |
|---|---|---|---|
| **Load components** | 40 residential and 3 commercial loads | 30 residential and 6 commercial loads | 15 residential and 3 commercial loads |
| **Total number of buildings** | 43 | 36 | 18 |

Also, to estimate the EV-charging load profiles, private EVs were selected to be of the model Nissan Leaf. It has a charging power capacity of 6.6 kW and a battery capacity of 40 kWh, which provides 270 kilometres of range. Furthermore, the utility EVs were chosen to be of the model LDV EV-80, which has a charging power capacity of 6.6 kW and a battery capacity of 56 kWh, which provides 190 kilometres of range [24].

A rule-based energy management strategy was additionally developed to decide the timing of EV charging – as part of the specifically developed rule-based, cycle-charging expert decision support system for the dispatch of the entire system described above. To coordinate the charging of EVs in MG 1, it was assumed that the charging of the 10 private EVs and 5 utility EVs occurs during the period between 9 p.m. and 5 a.m. using level-1 chargers. The reason behind the aforementioned scheduling strategy of EV-charging loads is to flatten the overall net load demand (total loads minus total renewable power generation) and improve the load factor, and consequently reduce the excess renewable power curtailments, as well as the size of the capital-intensive stationary battery bank – and, therefore, the total discounted system cost.

*4.4. Optimal sizing results*

Table 3 presents the optimal solution mix and the associated TNPCs of the three cases under consideration. As the table shows, the existence of WTs in MG 1 substantially contributes to reducing the optimal capacity of the PV system required and, therefore, the TNPC of the system. Another important observation lending further support to the above argument is that the TNPCs of MG 1 and MG 2a are comparable, in spite of the fact that the peak and average load on MG 1 are considerably greater than those on MG 2a. This provides another layer of evidence to the revealed insight that WTs play a significant role in decreasing the total estimated costs in MG 1, which can be primarily attributed to leveraging the underlying short-term and seasonal complementarities in solar PV and wind generations.

Table 3. Optimal mix of the candidate technologies and the corresponding TNPCs.

| Alg. | Sys. | PVs (no.) | BSS (no.) | WTs (no.) | TNPC ($) |
|---|---|---|---|---|---|
| MFOA | MG 1 | 102 | 11 | 3 | 435,539 |
|  | MG 2a | 784 | 7 | N/A* | 423,689 |
|  | MG 2b | 523 | 4 | N/A* | 266,754 |

*N/A = Not applicable because the WT is not considered as a candidate technology in MG 2a and MG 2b.

*4.5. Scenario analyses for MG 1: Indicative impact analyses of the timing of EV charging*

To evaluate the impact of the timing of EV charging on the costing and configuration of MGs, this section presents and discusses the results of scenario analyses carried out for MG 1. The considered scenarios include: (i) scheduling the charging of EVs to the late evening and early morning hours (9 p.m. to 5 a.m.), (ii) shifting the EV-charging loads to the afternoon hours (12 p.m. to 4 p.m.), and (iii) exclusion of EV-charging loads. For better-informed decision-making support, each of the scenarios, additionally, address two cases, namely: (1) without solar PV generation, and (2) without WT generation.

*5.4.1. Supplying the EV-charging loads in the late evening and early morning hours*

Table 4 presents the results of the optimum sizing of the components of MG 1, where the charging of EVs is scheduled to occur between the hours 9 p.m. to 5 a.m. It can be observed from the table that the exclusion of the solar PV from the candidate pool results in a slight decrease in the size of the BSS, but no significant change in the TNPC is expected compared to the baseline value. More specifically, a ~2% increase in the TNPC was observed, which equates to ~$9k. However, the case without a wind resource has had a more significant impact on the TNPC of the system. Specifically, a ~8% increase in the TNPC was observed (equating to ~$37k) due to the consequent modest increase in the size of solar PV panels and the BSS.

Table 4. Relative importance of solar PV and WT technologies on the economic viability of MG 1 with EV-charging loads supplied during the late evening and early morning hours.

| Scenario | PVs (no.) | BSS (no.) | WTs (no.) | TNPC ($) |
|---|---|---|---|---|
| Baseline | 102 | 11 | 3 | 435,539 |
| Without PV | - | 10 | 4 | 445,215 |
| Without WT | 614 | 14 | - | 472,087 |

*5.4.2. EV-charging loads shifted to the afternoon hours*

In this scenario, the EV-charging loads are shifted to the afternoon hours – specifically, from 12 p.m. to 4 p.m. – which has increased the TNPC by around 7% (equating to $32,156) compared to the original case where the EV-charging loads are served during the late evening and early morning hours (see Tables 4 and 5). The main underlying reason for this observation is the decreased load factor – defined as the average load divided by the peak load – which necessitates adding more solar PV capacity. Furthermore, the optimal capacity of the WT system and the battery bank is increased in the solar PV-less case, as one would expect, with a consequent TNPC increase of ~2% (equating to $6k). On the other hand, the WT-less scenario indicates a TNPC increase of ~5% (equating to $21k), which corroborates the relatively more significant impact of the WT system on the cost-effectiveness of MG 1 compared to the solar PV system.

Table 5. Relative importance of solar PV and WT technologies on the economic viability of MG 1 with EV-charging loads shifted to the afternoon hours.

| Scenario | PVs (no.) | BSS (no.) | WTs (no.) | TNPC ($) |
|---|---|---|---|---|
| Baseline | 397 | 6 | 3 | 467,695 |
| Without PV | - | 10 | 4 | 473,920 |
| Without WT | 667 | 13 | - | 489,093 |

*5.4.3. No EV-charging loads*

In this scenario, it is assumed that the transportation sector is not electrified. Expectedly, the TNPC is reduced compared to the other scenarios that consider the integration of EV-charging loads (see

Tables 4–6). Also, a comparison of the relevant solar PV-less and WT-less cases has provided another layer of evidence on the greater importance of WTs on the financial sustainability of the project compared to solar PV panels. More specifically, the PV-less case is associated with a total discounted system cost increase of ~3%, whereas the WT-less case is associated with a cost increase of ~8% compared to the associated baseline case.

Table 6. Relative importance of solar PV and WT technologies on the economic viability of MG 1 without EV-charging loads.

| Scenario | PVs (no.) | BSS (no.) | WTs (no.) | TNPC ($) |
| --- | --- | --- | --- | --- |
| Baseline | 211 | 11 | 2 | 426,284 |
| Without PV | - | 10 | 4 | 441,215 |
| Without WT | 614 | 14 | - | 468,087 |

## 5. Conclusions

MGs are typically associated with high capital and replacement costs, but low operation and maintenance costs. This makes the global (true) optimisation of the MG resources especially important – in the efforts to accelerate the transition to renewables. In particular, the integration of a high share of non-dispatchable and weather-dependent RESs, such as solar PV and wind, into MGs adds significant complexities to the conventional micro-power-system design problem. In addition to dealing with the high capital costs of RESs and the variability in their power outputs, there is a wide variety of operational- and planning-level constraints that need to be simultaneously met while optimally designing a MG.

Further compounding the MG design problem is the consideration of large-scale EV-charging loads, the appropriate coordination of supplying which can have a significant impact on the optimality of the MG design solutions. This is particularly salient in off-grid MG installations where there exists no connection to the wider utility grid to help in serving the peak load and/or absorbing the excess generation. Accordingly, there is a great risk of sub-optimality in non-grid-connected, EV-addressable MG designing problems, with potential overbuilt capacity or inadequate reliability implications. This necessitates the development of advanced meta-heuristic-based methods for the optimal sizing of stand-alone MGs. Such advanced meta-heuristics have a significant potential for providing improved observability of the entire search space of the MG design problem without the need for problem simplifications that might disconnect the underlying problem from reality.

In this context, this paper has presented a novel data-driven, metaheuristic-based off-grid MG capacity planning optimisation modelling framework which incorporates a coordinated EV-charging strategy. Based on the novel insights generated from the application of the proposed model to three conceptual off-grid MG cases in Aotea–Great Barrier Island, the importance of advanced meta-heuristic-based MG planning approaches that integrate specifically developed EV-charging-addressable dispatch strategies has been substantiated. The proposed method is readily scalable and adaptable for application to other off-grid MG cases. Also, the findings on the economic viability of optimally designed off-grid, 100%-renewable MGs are broadly relevant beyond the case studies considered – towards moving away from diesel dependency in isolated areas, which is essential to improved energy security.